\documentclass[preprint,prd,nofootinbib,onecolumn,showpacs,showkeys]{revtex4}%
\usepackage{amsfonts}
\usepackage{amsmath}
\usepackage{amssymb}
\usepackage{graphicx}
\usepackage{float}
\usepackage{rotating}%
\providecommand{\U}[1]{\protect\rule{.1in}{.1in}}

\ifx\pdfoutput\relax\let\pdfoutput=\undefined\fi
\newcount\msipdfoutput
\ifx\pdfoutput\undefined\else
\ifcase\pdfoutput\else
\ifx\paperwidth\undefined\else
\ifdim\paperheight=0pt\relax\else\pdfpageheight\paperheight\fi
\ifdim\paperwidth=0pt\relax\else\pdfpagewidth\paperwidth\fi
\fi\fi\fi
\begin{document}
\preprint{ }
\title[Intonation and Compensation]{Intonation and Compensation of Fretted String Instruments}
\author{Gabriele U. Varieschi}
\author{Christina M. Gower}
\affiliation{Department of Physics, Loyola Marymount University - Los Angeles, CA 90045,
USA\footnote{Email: gvarieschi@lmu.edu; cgower@lion.lmu.edu}}

\begin{abstract}
In this paper we present mathematical and physical models to be used in the
analysis of the problem of intonation of musical instruments such as guitars,
mandolins and the like, i.e., we study how to improve the tuning on these instruments.

This analysis begins by designing the placement of frets on the fingerboard
according to mathematical rules and the assumption of an ideal string, but
becomes more complicated when one includes the effects of deformation of the
string and inharmonicity due to other string characteristics. As a consequence
of these factors, perfect intonation of all the notes on the instrument can
never be achieved, but complex compensation procedures are introduced and
studied to minimize the problem.

To test the validity of these compensation procedures, we have performed
extensive measurements using standard monochord sonometers and other basic
acoustical devices, confirming the correctness of our theoretical models. In
particular, these experimental activities can be easily integrated into
standard acoustics courses and labs, and can become a more advanced version of
basic experiments with monochords and sonometers.

\end{abstract}
\startpage{1}
\endpage{ }
\maketitle
\tableofcontents

\section{\label{sect:introduction}Introduction}

The physics of musical instruments\ is a very interesting sub-field of
acoustics, which connects the mathematical models of vibrations and waves to
the world of art and musical performance. This connection between science and
music has always been present, since the origin of art and civilization.
Classic books on the field are for example \cite{Helmholtz}, \cite{Jeans},
\cite{Benade1}, \cite{Olson}. In the 6th century B.C., the mathematician and
philosopher Pythagoras was fascinated by music and by the intervals between
musical tones. He was probably the first to perform experimental studies of
the pitches of musical instruments and relate them to ratios of integer numbers.

This idea was the origin of the diatonic scale, which dominated much of
western music, and also of the so-called just intonation system which was used
for many centuries to tune musical instruments, based on perfect ratios of
whole numbers. Eventually, this system was abandoned in favor of a more
mathematically refined method for \textit{intonation} and tuning, the well
known equal temperament system, which was introduced by scholars such as
Vincenzo Galilei (Galileo's father), Marin Mersenne and Simon Stevin, in the
16th and 17th centuries, and also strongly advocated by musicians such as the
great J. S. Bach. In the equal-tempered scale, the interval of one octave is
divided into 12 equal sub-intervals (semitones), achieving a more uniform
intonation of musical instruments, especially when using all the 24 major and
minor keys, as in Bach's masterpiece, the \textquotedblleft Well Tempered
Clavier.\textquotedblright\ Historical discussion and complete reviews of the
different intonation systems can be found in Refs. \cite{Isacoff},
\cite{Barbour}, \cite{Loy}.

Mathematically, the twelve-tone equal temperament system requires the use of
irrational numbers, since for example the ratio of the frequencies of two
adjacent notes corresponds to $\sqrt[12]{2}$. On a \textit{fretted string
instrument} like a guitar, lute, mandolin, or similar, this intonation system
is accomplished by placing the frets along the fingerboard according to these
mathematical ratios. Unfortunately, even with the most accurate fret
placement, perfect instrument tuning is never achieved. This is due mainly to
the mechanical action of the player's fingers, which need to press the strings
down on the fingerboard while playing, thus altering their length, tension and
ultimately changing the frequency of the sound being produced. Other causes of
imperfect intonation include inharmonicity of the strings, due to their
intrinsic stiffness and other more subtle effects. An introduction to all
these effects can be found in Refs. \cite{Rossing} and \cite{Fletcher}.

Experienced luthiers and guitar manufacturers usually correct for this effect
by introducing the so-called \textit{compensation}, i.e., they slightly
increase the string length in order to compensate for the increased sound
frequency, resulting from the effects described above (see instrument building
techniques in \cite{Rodriguez}, \cite{Hopkin}, \cite{Middleton},
\cite{Cumpiano}, \cite{Lundberg}). Other solutions are reported in luthiers'
websites (\cite{gilbert}, \cite{byers2}, \cite{stenzel}, \cite{compensation},
\cite{doolin}) or in commercially patented devices (\cite{Jones},
\cite{Smith}, \cite{Feiten}). These empirical solutions can be improved by
studying the problem in a more scientific fashion, through proper modeling of
the string deformation and other effects, therefore leading to a new type of
fret placement which is more effective for the proper intonation of the instrument.

Some mathematical studies of the problem appeared in specialized journals for
luthiers and guitar builders (\cite{Bartolini}, \cite{Byers1}), but they were
particularly targeted to luthiers and manufacturers of a specific instrument
(typically classical guitar). We are not aware of similar scientific studies
being reported in physics or acoustics publications. For example, in general
physics journals we found only basic studies on guitar intonation and strings
(see \cite{1980AmJPh..48..362J}, \cite{1989PhTea..27..673H},
\cite{1991PhTea..29..438S}, \cite{1998AmJPh..66..144S},
\cite{0031-9120-38-4-303}, \cite{2006PhTea..44..465I},
\cite{2006PhTea..44..509L}, \cite{2007PhTea..45....4M},
\cite{2008PhTea..46..486L}), without any detailed analysis of the problem
outlined above.

Therefore, our objective is to review and improve the existing mathematical
models of compensation for fretted string instruments and to perform
experimental measures to test these models. In particular, the experimental
activities described in this paper were performed using standard lab equipment
(sonometers and other basic acoustic devices) in view of the pedagogical goal
of this project. In fact, all the experimental activities detailed in this
work can be easily introduced in standard sound and waves lab courses, as an
interesting variation of experiments usually performed with classic sonometers.

In the next section we will start by describing the geometry of the problem in
terms of a simple string deformation model. In Sect.
\ref{sect:compensation_model} we will examine the theoretical basis for the
compensation model being used, and in Sect.
\ref{sect:experimental_measurements} we will describe the outcomes of our
experimental activities.

\section{\label{sect:geometrical_model}Geometrical model of a fretted string}

In this section we will introduce the geometrical model of a guitar
fingerboard, review the practical laws for fret placement and study the
deformations of a \textquotedblleft fretted\textquotedblright\ string, i.e.,
when the string is pressed onto the fingerboard by the mechanical action of
the fingers.

\subsection{\label{sect:fret_placement}Fret placement on the fingerboard}

We start our analysis by recalling Mersenne's law which describes the
frequency $\nu$ of sound produced by a vibrating string \cite{Rossing},
\cite{Fletcher}:%

\begin{equation}
\nu_{n}=\frac{n}{2L}\sqrt{\frac{T}{\mu}}, \label{eqn2.1}%
\end{equation}
where $n=1$ refers to the fundamental frequency, while $n=2,3...$ to the
overtones. $L$ is the string length, $T$ is the tension, $\mu$ is the linear
mass density of the string (mass per unit length), and $v=\sqrt{T/\mu}$ is the
wave velocity.

In the equal-tempered musical scale an octave is divided into twelve
semitones, mathematically:%

\begin{equation}
\nu_{i}=\nu_{0}\ 2^{\frac{i}{12}}\simeq\nu_{0}\ (1.05946)^{i}, \label{eqn2.2}%
\end{equation}
where $\nu_{0}$ and $\nu_{i}$ are respectively the frequencies of the first
note in the octave and of the $i-th$ note ($i=1,2,...,12$). For $i=12$, we
obtain a frequency which is double that of the first note, as expected. Since
Mersenne's law states that the fundamental frequency of the vibrating string
is inversely proportional to the string length $L$, we simply combine Eqs.
(\ref{eqn2.1}) and (\ref{eqn2.2}) to determine the correct string lengths for
all the different notes ($i=1,2,3,...$) as a function of the original string
length $L_{0}$ (open string length, producing the first note of the octave
considered), assuming that the tension $T$ and the mass density $\mu$ are kept constant:%

\begin{equation}
L_{i}=L_{0}\ 2^{-\frac{i}{12}}\simeq L_{0}\ (0.943874)^{i}. \label{eqn2.3}%
\end{equation}
This equation can be immediately used to determine the fret placement on a
guitar or a similar instrument, since the frets essentially subdivide the
string length into the required sub-lengths.%

\begin{figure}
[ptb]
\begin{center}
\fbox{\ifcase\msipdfoutput
\includegraphics[
width=\textwidth 
]%
{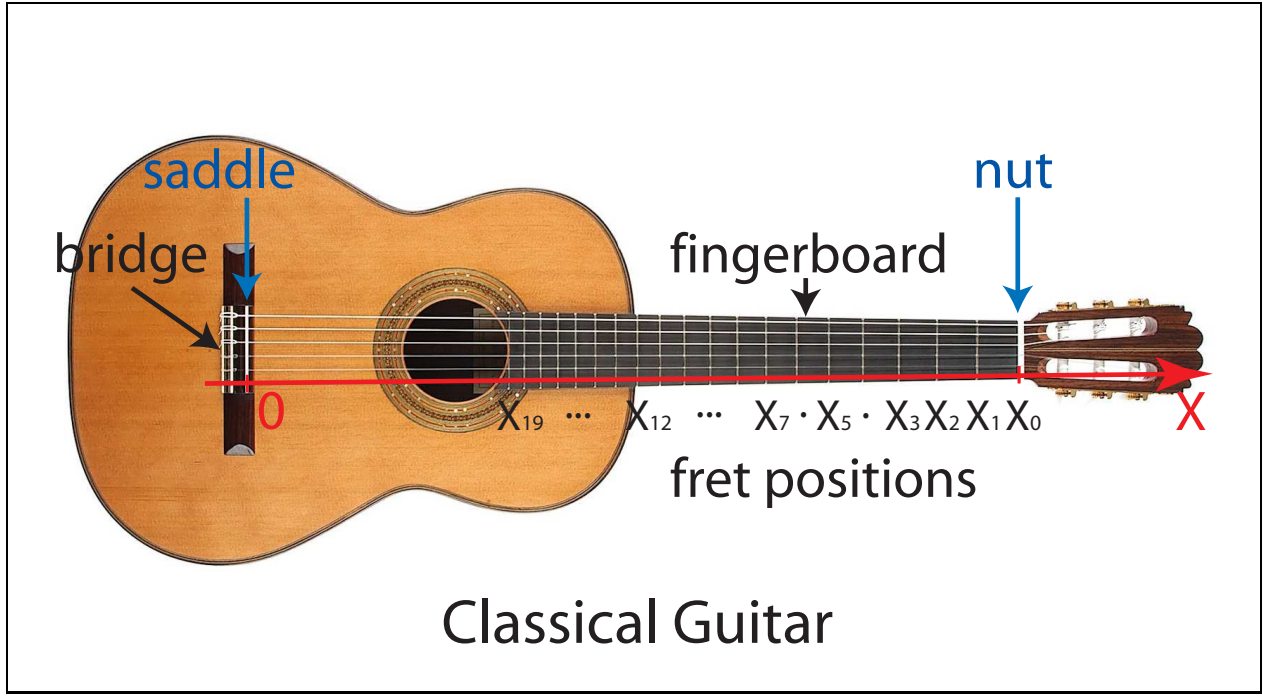}%
\else
\includegraphics[
width=\textwidth
]%
{C:/swp55/Docs/guitar1.1/AJP/arxiv_v2/graphics/Varieschi_GowerFig01__1.pdf}%
\fi
}\caption[Illustration of a classical guitar showing our coordinate
system.]{Illustration of a classical guitar showing our coordinate system,
from the saddle toward the nut, used to measure the fret positions on the
fingerboard (guitar by Michael Peters - photo by Trilogy Guitars, reproduced
with permission).}%
\label{fig1}%
\end{center}
\end{figure}
In Figure \ref{fig1} we show a picture of a classical guitar as a reference.
The string length is the distance between the saddle\footnote{The saddle is
the white piece of plastic or other material located near the bridge, on which
the strings are resting. The strings are usually attached to the bridge, which
is located on the left of the saddle. On other type of guitars, or other
fretted instruments, the strings are attached directly to the bridge (without
using any saddle). In this case the string length would be the distance
between the bridge and the nut. Our analysis would not be different in this
case: the bridge position would simply replace the saddle position.} and the
nut, while the frets are placed on the fingerboard at appropriate distances.
We prefer to use the coordinate $X$, as illustrated in the same figure, to
denote the position of the frets, measured from the saddle toward the nut
position. $X_{0}$ will denote the position of the nut (the \textquotedblleft
zero\textquotedblright\ fret), while $X_{i}$, $i=1,2,...$, are the positions
of the frets of the instrument. On a classical guitar there are usually up to
$19-20$ frets on the fingerboard and they are realized by inserting thin
pieces of a special metal wire in the fingerboard, so that the frets will rise
about $1.0-1.5\ mm$ above the fingerboard level.

The positioning of the frets follows Eq. (\ref{eqn2.3}), which we rewrite in
terms of our new variable $X$:%

\begin{equation}
X_{i}=X_{0}\ 2^{-\frac{i}{12}}\simeq X_{0}\ (0.943874)^{i}\simeq
X_{0}\ \left(  \frac{17}{18}\right)  ^{i}, \label{eqn2.4}%
\end{equation}
where the last approximation in the previous equation is the one historically
employed by luthiers to practically locate the fret positions. This is usually
called the \textquotedblleft rule of 18,\textquotedblright\ which requires
placing the first fret at a distance from the nut corresponding to $\frac
{1}{18}$ of the string length (or $\frac{17}{18}$ from the saddle); then place
the second fret at a distance from the first fret corresponding to $\frac
{1}{18}$ of the remaining length between the first fret and the saddle, and so
on. Since $\frac{17}{18}=0.944444\simeq0.943874$, this empirical method is
usually accurate enough for practical fret placement\footnote{Following Eq.
(\ref{eqn2.4}), frets number $5$, $7$, $12$, and $19$, are particularly
important since they (approximately) correspond to vibrating string lengths
which are respectively $3/4$, $2/3$, $1/2$, and $1/3$ of the full length, in
line with the Pythagorean original theory of monochords.}, although modern
luthiers use fret placement templates based on the decimal expression in Eq.
(\ref{eqn2.4}).

\subsection{\label{sect:deformation_model}Deformation model of a fretted
string}

\begin{figure}
[ptb]
\begin{center}
\fbox{\ifcase\msipdfoutput
\includegraphics[
width=\textwidth
]%
{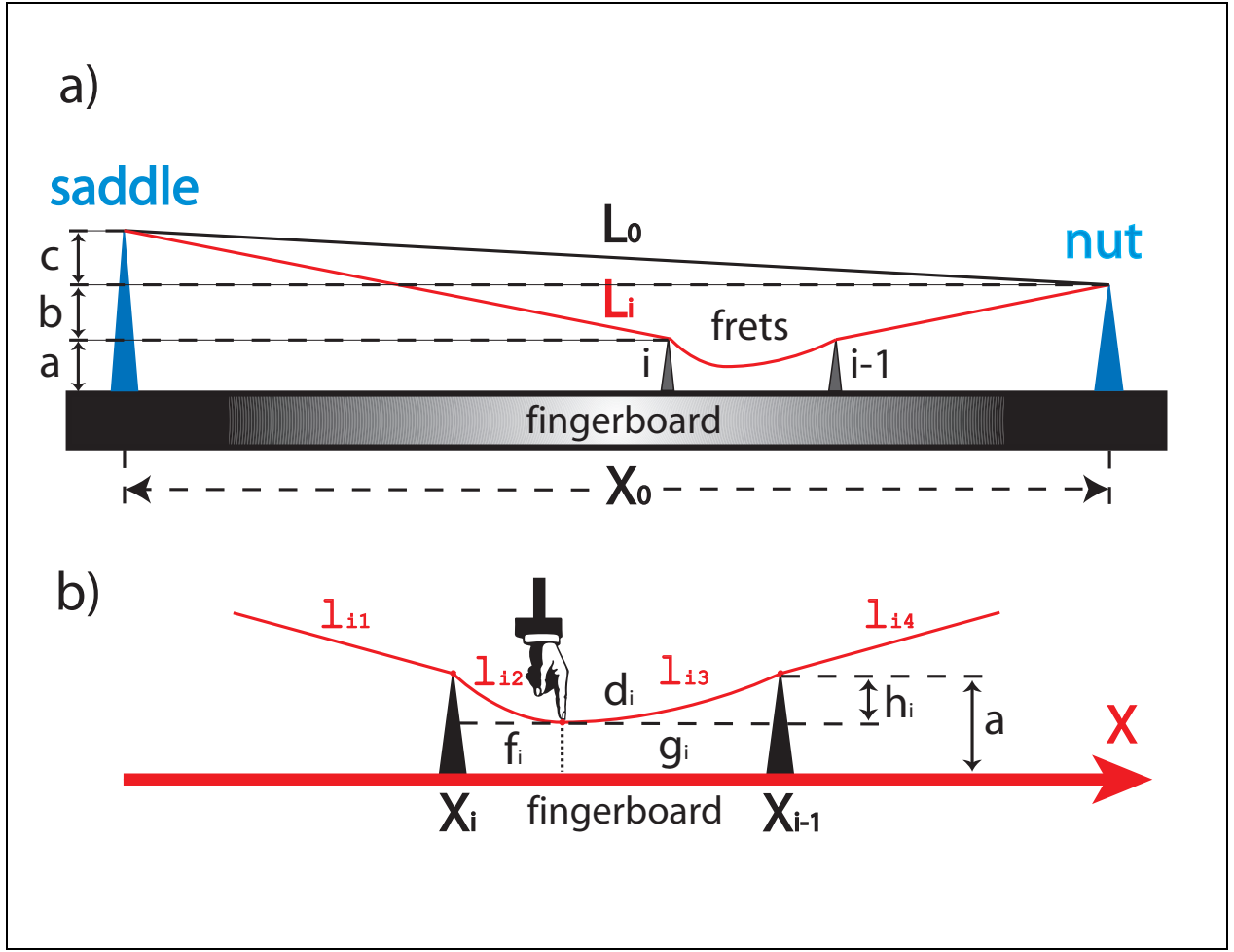}%
\else
\includegraphics[
width=\textwidth
]%
{C:/swp55/Docs/guitar1.1/AJP/arxiv_v2/graphics/Varieschi_GowerFig02__2.pdf}%
\fi
}\caption[Geometrical deformation model of a guitar string.]{Geometrical
deformation model of a guitar string. In part a) we show the original string
in black (of length $L_{0}$) and the deformed string in red (of length $L_{i}%
$) when it is pressed between frets $i$ and $i-1$. In part b) we show the
details of our deformation model, in terms of the four different sub-lengths
$l_{i1}-l_{i4}$ of the deformed string.}%
\label{fig2}%
\end{center}
\end{figure}

Figure \ref{fig2} illustrates the geometrical model of a fretted\ string,
i.e., when a player's finger or other device is pressing the string down to
the fingerboard, until the string is resting on the desired $i-th$ fret, thus
producing the $i-th$ note when the string is plucked. In this figure we use a
notation similar to the one developed in Refs. \cite{Bartolini},
\cite{Byers1}, but we will introduce a different deformation model.%

Figure \ref{fig2}a shows the general geometrical variables for a guitar
string. The distance $X_{0}$ between the saddle and the nut is also called the
scale-length of the guitar (typically between $640-660\ mm$ for a modern
classical guitar) but this is not exactly the same as the real string length
$L_{0}$, because saddle and nut usually have slightly different heights above
the fingerboard surface. The connection between $L_{0}$ and $X_{0}$ is simply:%

\begin{equation}
L_{0}=\sqrt{X_{0}^{2}+c^{2}}. \label{eqn2.5}%
\end{equation}

The metal frets rise above the fingerboard by a distance $a$ as shown in
Figure \ref{fig2}. The heights of the nut and saddle above the top of the
frets are labeled in Figure \ref{fig2} as $b$ and $c$, respectively. All these
heights are greatly exaggerated; they are usually small compared to the string
length. The standard fret positions are again denoted by $X_{i}$ and, in
particular, we show the situation where the string is pressed between frets
$i$ and $i-1$, thus reducing the vibrating portion of the string to the part
between the saddle and the $i-th$ fret.

Figure \ref{fig2}b shows the details of the deformation caused by the action
of a finger between the two frets. Previous works (\cite{Bartolini},
\cite{Byers1}) modeled this shape simply as a sort of \textquotedblleft
knife-edge\textquotedblright\ deformation which is not quite comparable to the
action of a fingertip. We improved on this point by assuming a more
\textquotedblleft rounded\textquotedblright\ deformation, considering a curved
shape as in Figure \ref{fig2}b. The action of the finger depresses the string
behind the $i-th$ fret by an amount $h_{i}$ below the fret level (not
necessarily corresponding to the full height $a$) and at a distance $f_{i}$,
compared to the distance $d_{i}$ between consecutive frets.

In Sect. \ref{sect:experimental_measurements} we will describe how to set all
these parameters to the desired values with our experimental device and
simulate all possible deformations of the string. It is necessary for our
compensation model, described in the next section, to compute exactly the
length of the deformed string for any fret value $i$. As shown in Fig.
\ref{fig2}, the deformed length $L_{i}$ of the entire string is the sum of the
lengths of the four different parts:%

\begin{equation}
L_{i}=l_{i1}+l_{i2}+l_{i3}+l_{i4}, \label{eqn2.6}%
\end{equation}
where these four sub-lengths can be evaluated from the geometrical parameters
as follows:%

\begin{align}
l_{i1}  &  =\sqrt{\left(  X_{0}\ 2^{-\frac{i}{12}}\right)  ^{2}+(b+c)^{2}%
}\label{eqn2.7}\\
l_{i2}  &  =h_{i}\sqrt{1+\frac{f_{i}^{2}}{4h_{i}^{2}}}+\frac{f_{i}^{2}}%
{4h_{i}}\ln\left[  \frac{2h_{i}}{f_{i}}\left(  1+\sqrt{1+f_{i}^{2}/4h_{i}^{2}%
}\right)  \right] \nonumber\\
l_{i3}  &  =h_{i}\sqrt{1+\frac{g_{i}^{2}}{4h_{i}^{2}}}+\frac{g_{i}^{2}}%
{4h_{i}}\ln\left[  \frac{2h_{i}}{g_{i}}\left(  1+\sqrt{1+g_{i}^{2}/4h_{i}^{2}%
}\right)  \right] \nonumber\\
l_{i4}  &  =\sqrt{X_{0}^{2}\left(  1-2^{-\frac{i-1}{12}}\right)  ^{2}+b^{2}%
}.\nonumber
\end{align}

In Eq. (\ref{eqn2.7}) the sub-lengths $l_{i2}$ and $l_{i3}$ were obtained by
using a simple parabolic shape for the \textquotedblleft
rounded\textquotedblright\ deformation shown in Fig. \ref{fig2}b, due to the
action of the player's fingertip. They were computed by integrating the length
of the two parabolic arcs shown in Fig. \ref{fig2}b, in terms of the distances
$f_{i}$, $g_{i}$ and $h_{i}$.

The distances between consecutive frets are calculated as:%

\begin{equation}
d_{i}=f_{i}+g_{i}=X_{i-1}-X_{i}=X_{0}\ 2^{-\frac{i}{12}}\left(  2^{\frac
{1}{12}}-1\right)  \label{eqn2.8}%
\end{equation}
so that, given the values of $X_{0}$, $a$, $b$, $c$, $h_{i}$ and $f_{i}$, we
can compute for any fret number $i$ the values of all the other quantities and
the deformed length $L_{i}$. We will see in the next section that the
fundamental geometrical quantities of the compensation model are defined as:%

\begin{equation}
Q_{i}=\frac{L_{i}-L_{0}}{L_{0}} \label{eqn2.9}%
\end{equation}
and they can also be computed for any fret $i$ using Eqs. (\ref{eqn2.5}) -
(\ref{eqn2.8}).

\section{\label{sect:compensation_model}Compensation model}

In this section we will describe the model used to compensate for the string
deformation and for the inharmonicity of a vibrating string, basing our
analysis on the work done by G. Byers (\cite{byers2}, \cite{Byers1}).

\subsection{\label{sect:vibrations}Vibrations of a stiff string}

Strings used in musical instruments are not perfectly elastic, but possess a
certain amount of \textquotedblleft stiffness\textquotedblright\ or
inharmonicity which affects the frequency of the sound produced. Mersenne's
law in Eq. (\ref{eqn2.1}) needs to be modified to include this property of
real strings, yielding the following result (see Ref. \cite{Morse}, chapter 4,
section 16):%

\begin{equation}
\nu_{n}\simeq\frac{n}{2L}\sqrt{\frac{T}{\rho S}}\left[  1+\frac{2}{L}%
\sqrt{\frac{ESk^{2}}{T}}+\left(  4+\frac{n^{2}\pi^{2}}{2}\right)
\frac{ESk^{2}}{TL^{2}}\right]  , \label{eqn3.1}%
\end{equation}
where we have rewritten the linear mass density of the string as $\mu=\rho S$
($\rho$ is the string density and $S$ the cross section area). The correction
terms inside the square brackets are due to the string stiffness and related
to the modulus of elasticity (or Young's modulus) $E$ and to the radius of
gyration $k$ (equal to the string radius divided by two, for a simple unwound
steel or nylon string). Following Ref. \cite{Morse}, we will use c.g.s. units
in the rest of the paper and in all computations, except when quoting some
geometrical parameters for which it will be more convenient to use millimeters.

The previous equation is an approximation valid for $\frac{ESk^{2}}{TL^{2}%
}<\frac{1}{n^{2}\pi^{2}}$, a condition which is usually met in practical
situations\footnote{The condition is equivalent to $n^{2}<\frac{1}{\pi^{2}%
}\frac{TL^{2}}{ESk^{2}}\approx369;\ 803;\ 5052$, where the numerical values
are related to the three steel strings we will use in Sect.
\ref{sect:experimental_measurements} (see string properties in Table
\ref{TableKey:table1}) and for the shortest possible vibrating length
$L\simeq\frac{1}{3}L_{0}\simeq21.5\ cm$. The approximation in Eq.
(\ref{eqn3.1}) is certainly valid for our strings, for at least $n\lesssim
19$.}. When the stiffness factor $\frac{ESk^{2}}{TL^{2}}$ is negligible, Eq.
(\ref{eqn3.1}) reduces to the original Eq. (\ref{eqn2.1}). On the contrary,
when this factor increases and becomes important, the allowed frequencies also
increase, following the last equation, and the overtones ($n=2,3,...$)
increase in frequency more rapidly than the fundamental tone ($n=1$). The
sound produced is no longer \textquotedblleft harmonic\textquotedblright%
\ since the overtone frequencies are no longer simple multiples of the
fundamental one, as seen from Eq. (\ref{eqn3.1}). In addition, the deformation
of the fretted string, described in the previous section, will alter the
string length $L$ and, as a consequence of this effect, will also change the
tension $T$ and the cross section $S$ in the last equation. These are the main
causes of the intonation problem being studied. Additional causes that we
cannot address in this work are the imperfections of the strings (non uniform
cross section or density), the motion of the end supports (especially the
saddle and the bridge, transmitting the vibrations to the rest of the
instrument) which also changes the string length, the effects of friction, and others.

Following Byers \cite{Byers1} we define $\alpha_{n}=\left(  4+\frac{n^{2}%
\pi^{2}}{2}\right)  $ and $\beta=\sqrt{\frac{ESk^{2}}{T}}$, so that we can
simplify Eq. (\ref{eqn3.1}):%

\begin{equation}
\nu_{n}\simeq\frac{n}{2L}\sqrt{\frac{T}{\rho S}}\left[  1+2\frac{\beta}%
{L}+\alpha_{n}\frac{\beta^{2}}{L^{2}}\right]  . \label{eqn3.2}%
\end{equation}
We then consider just the fundamental tone ($n=1$) as being the frequency of
the sound perceived by the human ear\footnote{This statement is also an
approximation since the pitch (or perceived frequency) is affected by the
presence of the overtones. See for example the discussion of the psychological
characteristics of music in Olson \cite{Olson}.}:%

\begin{equation}
\nu_{1}\simeq\frac{1}{2L}\sqrt{\frac{T}{\rho S}}\left[  1+2\frac{\beta}%
{L}+\alpha\frac{\beta^{2}}{L^{2}}\right]  , \label{eqn3.3}%
\end{equation}
where $\alpha=\alpha_{1}=\left(  4+\frac{\pi^{2}}{2}\right)  $ and $\beta$ is
defined as above. In Eq. (\ref{eqn3.3}) $L$ represents the vibrating length of
the string, which in our case is the length $l_{i1}$ when the string is
pressed down onto the $i-th$ fret. To further complicate the problem, the
quantities $T$, $S$ and $\beta$ in Eq. (\ref{eqn3.3}) depend on the actual
total length of the string $L_{i}$, as computed in Eq. (\ref{eqn2.6}). In
other words, we tune the open string, of original length $L_{0}$, at the
appropriate tension $T$, but when the string is \textquotedblleft
fretted\textquotedblright\ its length is changed from $L_{0}$ to $L_{i}$, thus
slightly altering the tension, the cross section, and also $\beta$ which is a
function of the previous two quantities. This is the origin of the lack of
intonation, common to all fretted instruments, which calls for an appropriate
compensation mechanism, which will be analyzed in the next section.

\subsection{\label{sect:compensation_nut_saddle}Compensation at nut and
saddle}

The proposed solution \cite{Byers1} to the intonation problem is to adjust the
fret positions to accommodate for the frequency changes described in the
previous equation. The vibrating lengths $l_{i1}$ are recomputed as
$l_{i1}^{\prime}=l_{i1}+\Delta l_{i1}$, where $\Delta l_{i1}$ represents a
small adjustment in the placement of the frets, so that the fundamental
frequency from Eq. (\ref{eqn3.3}) will match the ideal frequency of Eq.
(\ref{eqn2.2}) and the fretted note will be in tune.

The ideal frequency $\nu_{i}$ of the $i-th$ note can be expressed by combining
together Eqs. (\ref{eqn2.2}) and (\ref{eqn3.3}):%

\begin{equation}
\nu_{i}=\nu_{0}\ 2^{\frac{i}{12}}\simeq\frac{1}{2L_{0}}\sqrt{\frac{T(L_{0}%
)}{\rho S(L_{0})}}\left[  1+2\frac{\beta(L_{0})}{L_{0}}+\alpha\frac{\left[
\beta(L_{0})\right]  ^{2}}{L_{0}^{2}}\right]  \ 2^{\frac{i}{12}},
\label{eqn3.3.1}%
\end{equation}
where all the quantities on the right-hand side of the previous equation are
related to the open string length $L_{0}$, since $\nu_{0}$ is the frequency of
the open string note. On the other hand, we can write the same frequency
$\nu_{i}$ using Eq. (\ref{eqn3.3}) directly for the fretted note:%

\begin{equation}
\nu_{i}\simeq\frac{1}{2l_{i1}^{\prime}}\sqrt{\frac{T(L_{i})}{\rho S(L_{i})}%
}\left[  1+2\frac{\beta(L_{i})}{l_{i1}^{\prime}}+\alpha\frac{\left[
\beta(L_{i})\right]  ^{2}}{l_{i1}^{\prime2}}\right]  , \label{eqn3.3.2}%
\end{equation}
where now we use the \textquotedblleft adjusted\textquotedblright\ vibrating
length $l_{i1}^{\prime}$ for the fretted note and all the other quantities on
the right-hand side of Eq. (\ref{eqn3.3.2}) depend on the fretted string
length $L_{i}$. By comparing Eqs. (\ref{eqn3.3.1}) and (\ref{eqn3.3.2}) we
obtain the master equation for our compensation model:%

\begin{equation}
\frac{1}{2L_{0}}\sqrt{\frac{T(L_{0})}{\rho S(L_{0})}}\left[  1+2\frac
{\beta(L_{0})}{L_{0}}+\alpha\frac{\left[  \beta(L_{0})\right]  ^{2}}{L_{0}%
^{2}}\right]  \ 2^{\frac{i}{12}}=\frac{1}{2l_{i1}^{\prime}}\sqrt{\frac
{T(L_{i})}{\rho S(L_{i})}}\left[  1+2\frac{\beta(L_{i})}{l_{i1}^{\prime}%
}+\alpha\frac{\left[  \beta(L_{i})\right]  ^{2}}{l_{i1}^{\prime2}}\right]  .
\label{eqn3.4}%
\end{equation}

We obtained an approximate solution\footnote{Our solution in Eq.
(\ref{eqn3.5}) differs from the similar solution obtained by Byers et al. (Eq.
17 in Ref. \cite{Byers1}). We believe that this is due to a minor error in
their computation, which yields only minimal changes in the numerical results.
Therefore, the compensation procedure used by G. Byers in his guitars is
essentially correct and practically very effective in improving the intonation
of his instruments.} of the previous equation by Taylor expanding the
right-hand side in terms of $\Delta l_{i1}$ and by solving the resulting
expression for the new vibrating lengths $l_{i1}^{\prime}$:%

\begin{equation}
l_{i1}^{\prime}\simeq l_{i1}\left\{  1+\frac{\left[  1+\frac{2\beta(L_{0}%
)}{l_{i1}}+\frac{\alpha\left[  \beta(L_{0})\right]  ^{2}}{l_{i1}^{2}}\right]
-\frac{1}{\left[  1+Q_{i}(1+R)\right]  }\left[  1+\frac{2\beta(L_{0})}{L_{0}%
}+\frac{\alpha\left[  \beta(L_{0})\right]  ^{2}}{L_{0}^{2}}\right]  }{\left[
1+\frac{4\beta(L_{0})}{l_{i1}}+\frac{3\alpha\left[  \beta(L_{0})\right]  ^{2}%
}{l_{i1}^{2}}\right]  }\right\}  . \label{eqn3.5}%
\end{equation}
In this equation the quantities $Q_{i}$ are derived from Eq. (\ref{eqn2.9})
and from our new deformation model described in Sect.
\ref{sect:deformation_model}, while an additional experimental quantity $R$ is
introduced in the previous equation and defined as (see Ref. \cite{Byers1} for details):%

\begin{equation}
R=\left[  \frac{d\nu}{dL}\right]  _{L_{0}}\frac{L_{0}}{\nu_{0}},
\label{eqn3.6}%
\end{equation}
i.e., the frequency change $d\nu$ relative to the original frequency $\nu_{0}%
$, induced by an infinitesimal string length change $dL$, relative to the
original string length $L_{0}$. This quantity will be measured in Sect.
\ref{sect:experimental_measurements} for the strings we used in this project.

The new vibrating lengths $l_{i1}^{\prime}$ from Eq. (\ref{eqn3.5}) correspond
to new fret positions $X_{i}^{\prime}$, since $X_{i}^{\prime}=\sqrt
{l_{i1}^{\prime2}-(b+c)^{2}}\simeq l_{i1}^{\prime}$ for $(b+c)\ll
l_{i1}^{\prime}$. A similar relation also holds between $X_{i}$ and $l_{i1}$
(see Fig. \ref{fig2}) so that the same Eq. (\ref{eqn3.5}) can be used to
determine the new fret positions from the old ones:%

\begin{equation}
X_{i}^{\prime}\simeq X_{i}\left\{  1+\frac{\left[  1+\frac{2\beta(L_{0}%
)}{l_{i1}}+\frac{\alpha\left[  \beta(L_{0})\right]  ^{2}}{l_{i1}^{2}}\right]
-\frac{1}{\left[  1+Q_{i}(1+R)\right]  }\left[  1+\frac{2\beta(L_{0})}{L_{0}%
}+\frac{\alpha\left[  \beta(L_{0})\right]  ^{2}}{L_{0}^{2}}\right]  }{\left[
1+\frac{4\beta(L_{0})}{l_{i1}}+\frac{3\alpha\left[  \beta(L_{0})\right]  ^{2}%
}{l_{i1}^{2}}\right]  }\right\}  . \label{eqn3.7}%
\end{equation}

At this point a luthier should position the frets on the fingerboard according
to Eq. (\ref{eqn3.7}) which is not anymore in the canonical form of the
original Eq. (\ref{eqn2.4}). Moreover, each string would get slightly
different fret positions, since the physical properties such as tension, cross
section, etc., are different for the various strings of a musical instrument.
Therefore, this compensation solution would be very difficult to be
implemented practically and would also affect the playability of the
instrument\footnote{Nevertheless some luthiers actually construct guitars
where the individual frets under each string are adjustable in position by
moving them slightly along the fingerboard. Each note of the guitar is then
individually fine-tuned to achieve the desired intonation, requiring a very
time consuming tuning procedure.}.

An appropriate compromise, also introduced by Byers \cite{Byers1}, is to fit
the new fret positions $\left\{  X_{i}^{\prime}\right\}  _{i=1,2,...}$ to a
canonical fret position equation (similar to the original Eq. (\ref{eqn2.4}))
of the form:%

\begin{equation}
X_{i}^{\prime}=X_{0}^{\prime}\ 2^{-\frac{i}{12}}+\Delta S \label{eqn3.8}%
\end{equation}
where $X_{0}^{\prime}$ is a new scale length for the string and $\Delta S$ is
the \textquotedblleft saddle setback,\textquotedblright\ i.e., the distance by
which the saddle position should be shifted from its original position
(usually $\Delta S>0$ and the saddle is moved away from the nut). The nut
position is also shifted, but we require to keep the string scale at the
original value $X_{0}$, therefore we need $X_{NUT}^{\prime}+\Delta S=X_{0}$,
where $X_{NUT}^{\prime}$ is the new nut position in the primed coordinates.
Introducing the shift in the nut position $\Delta N$ as $X_{NUT}^{\prime
}=X_{0}^{\prime}+\Delta N$ and combining together the last two equations, we
obtain the definition of the \textquotedblleft nut
adjustment\textquotedblright\ $\Delta N$ as:%

\begin{equation}
\Delta N=X_{0}-\left(  X_{0}^{\prime}+\Delta S\right)  . \label{eqn3.9}%
\end{equation}
This is typically a negative quantity, indicating that the nut has to be moved
slightly forward toward the saddle.

Finally, instead of adopting a new scale length $X_{0}^{\prime}$, the luthier
might want to keep the same original scale length $X_{0}$ and keep the fret
positions according to the original Eq. (\ref{eqn2.4}). Since the corrections
and the effects we described above are essentially all linear with respect to
the scale length adopted, it will be sufficient to rescale the nut and saddle
adjustment as follows:%

\begin{align}
\Delta S_{resc}  &  =\frac{X_{0}}{X_{0}^{\prime}}\Delta S\label{eqn3.10}\\
\Delta N_{resc}  &  =\frac{X_{0}}{X_{0}^{\prime}}\Delta N.\nonumber
\end{align}

This final rescaling is also practically needed on a guitar or other fretted
instrument, because the compensation procedure described above has to be
carried out independently on each string of the instrument, i.e., all the
quantities in the equations of this sections should be rewritten adding a
string index $j=1,2,...,6$, for the six guitar strings. Each string would get
a particular saddle and nut correction, but once these corrections are all
rescaled according to Eq. (\ref{eqn3.10}) the luthier can still set the frets
according to the original Eq. (\ref{eqn2.4}). The saddle and nut will be
shaped in a way to incorporate all the saddle-nut compensation adjustments for
each string of the instrument (see \cite{Byers1}, \cite{byers2} for practical
illustrations of these techniques).

In practice, this compensation procedure does not change the original fret
placement and the scale length of the guitar, but requires very precise nut
and saddle adjustments for each of the strings of the instrument, using Eq.
(\ref{eqn3.10}). Again, this is just a convenient approximation of the full
compensation procedure, which would require repositioning all frets according
to Eq. (\ref{eqn3.7}), but this would not be a very practical solution.

In the next section we will detail the experimental measures we performed
following the deformation and compensation models outlined above. Since all
our measurements will be carried out using a monochord apparatus, we will work
essentially with a single string and not a whole set of six strings, as in a
real guitar. Therefore, we will use all the equations outlined above without
adding the additional string index $j$. However, it would be easy to modify
our discussion in order to extend the deformation-compensation model to the
case of a multi-string apparatus.

\section{\label{sect:experimental_measurements}Experimental measurements}

In Figure \ref{fig3} we show the experimental setup we used for our
measurements. Since our goal was simply to test the physics involved in the
intonation problem and not to build musical instruments or improve their
construction techniques, we used standard lab equipment, as illustrated in the figure.

\subsection{\label{description_apparatus}Description of the apparatus}

\begin{figure}
[ptb]
\begin{center}
\fbox{\ifcase\msipdfoutput
\includegraphics[
width=\textwidth
]%
{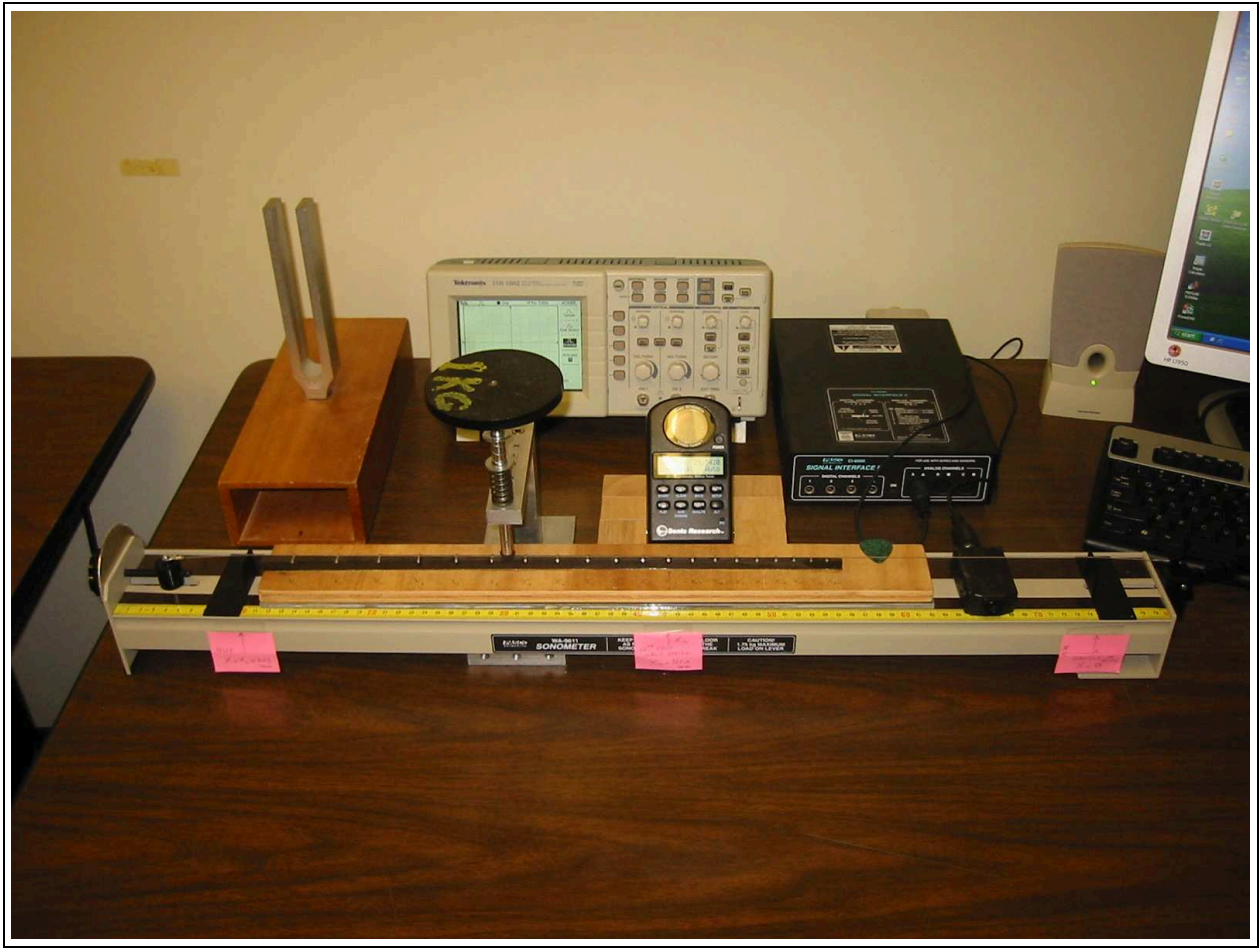}%
\else
\includegraphics[
width=\textwidth
]%
{C:/swp55/Docs/guitar1.1/AJP/arxiv_v2/graphics/Varieschi_GowerFig03__3}%
\fi
}\caption[Our experimental apparatus is composed of a standard sonometer to
which we added a classical guitar fingerboard.]{Our experimental apparatus is
composed of a standard sonometer to which we added a classical guitar
fingerboard. Also shown is a mechanical device used to press the string on the
fingerboard and several different instruments used to measure sound
frequencies.}%
\label{fig3}%
\end{center}
\end{figure}

A standard PASCO sonometer WA-9613 \cite{Sonometer1} was used as the main
device for the experiment. This apparatus includes a set of steel strings of
known linear density and diameter, and two adjustable bridges which can be
used to simulate nut and saddle of a guitar. The string tension can be
measured by using the sonometer tensioning lever, or adjusted directly with
the string tensioning screw (on the left of the sonometer, as seen in Fig.
\ref{fig3}). In particular, this adjustment allowed the direct measurement for
each string of the $R$ parameter in Eq. (\ref{eqn3.6}), by slightly stretching
the string and measuring the corresponding frequency change.

On top of the sonometer we placed a piece of a classical guitar fingerboard
with scale length $X_{0}=645\ mm$ (visible in Fig. \ref{fig3} as a thin black
object\ with twenty metallic frets, glued to a wooden board to raise it almost
to the level of the string). The geometrical parameters in Fig. \ref{fig2}
were set as follows: $a=1.3\ mm$ (fret thickness), $b=1.5\ mm$, $c=0.0\ mm$
(since we used two identical sonometer bridges as nut and saddle). This
arrangement ensured that the metal strings produced a good quality sound,
without \textquotedblleft buzzing\textquotedblright\ or producing undesired
noise, when the sonometer was \textquotedblleft played\textquotedblright\ like
a guitar by gently plucking the string. Also, since we set $c=0$, the open
string length is equal to the scale length: $L_{0}=X_{0}=645\ mm$.

The mechanical action of the player's finger pressing on the string was
produced by using a spring loaded device (also shown in Fig. \ref{fig3},
pressing between the sixth and seventh fret) with a rounded end to obtain the
deformation model illustrated in Fig \ref{fig2}b. Although we tried different
possible ways of pressing on the strings, for the measurements described in
this section we were always pressing halfway between the frets ($f_{i}%
=g_{i}=\frac{1}{2}d_{i}$) and all the way down on the fingerboard
($h_{i}=a=1.3\ mm$). In this way, all the geometrical parameters of Fig.
\ref{fig2} were defined and the fundamental quantities $Q_{i}$ of Eq.
(\ref{eqn2.9}) could be computed.%

\bigskip

The sound produced by the plucked string (which was easily audible, due to the
resonant body of the sonometer) was analyzed with different devices, in order
to accurately measure its frequency. At first we used the sonometer detector
coil or a microphone, connected to a digital oscilloscope, or alternatively to
a computer through a digital signal interface, as shown also in Fig.
\ref{fig3}. All these devices could measure frequencies in an accurate way,
but we decided to use for most of our measurements a professional digital
tuner \cite{turbotuner}, which could discriminate frequencies at the level of
$\pm0.1\ cents$ \footnote{The $cent$ is a logarithmic unit of measure used for
musical intervals. The octave is divided into twelve semitones, each of which
is subdivided in $100\ cents$, thus the octave is divided into $1200\ cents$.
Since an octave corresponds to a frequency ratio of $2:1$, one cent is
precisely equal to an interval of $2^{\frac{1}{1200}}$. Given two frequencies
$a$ and $b$ of two different notes, the number $n$ of cents between the notes
is $n=1200\log_{2}(a/b)\simeq3986\log_{10}(a/b)$. Alternatively, given a note
$b$ and the number $n$ of cents in the interval, the second note $a$ of the
interval is $a=b\times2^{\frac{n}{1200}}$.}. This device is shown near the
center of Fig. \ref{fig3}, just behind the sonometer.

\subsection{\label{string_properties}String properties and experimental
results}

For our experimental tests we chose three of the six steel guitar strings
included with the PASCO sonometer. Their physical characteristics and the
compensation parameters are described in Table \ref{TableKey:table1}.%

\begin{table}[tbp] \centering
\begin{tabular}
[c]{||c|c|c|c||}\hline\hline
\textbf{\#} & \textbf{String 1} & \textbf{String 2} & \textbf{String
3}\\\hline
Open string note & $C_{3}$ & $F_{3}$ & $C_{4}$\\\hline
Open string frequency ($Hz$) & $130.813$ & $174.614$ & $261.626$\\\hline
Radius ($cm$) & $0.0254$ & $0.0216$ & $0.0127$\\\hline
Linear density $\mu$ ($g/cm$) & $0.0150$ & $0.0112$ & $0.0039$\\\hline
Tension ($dyne$) & $5.16\times10^{6}$ & $5.88\times10^{6}$ & $4.41\times
10^{6}$\\\hline
Young's modulus $E$ ($dyne/cm^{2}$) & $2.00\times10^{12}$ & $2.00\times
10^{12}$ & $2.00\times10^{12}$\\\hline
R parameter & $130$ & $199$ & $78.7$\\\hline
Rescaled saddle setback $\Delta S_{resc}$ ($cm$) & $0.733$ & $0.998$ &
$0.518$\\\hline
Rescaled nut adjustment $\Delta N_{resc}$ ($cm$) & $-2.31$ & $-2.41$ &
$-1.35$\\\hline\hline
\end{tabular}
\caption{The physical characteristics and the compensation parameters for the three steel strings used in our experimental tests are summarized here.}\label{TableKey:table1}%
\end{table}%

The open string notes and related frequencies were chosen so that the sound
produced using all the twenty frets of our fingerboard would span over 2-3
octaves, and the tensions were set accordingly. We used a value for Young's
modulus which is typical of steel strings and we measured the $R$ parameter in
Eq. (\ref{eqn3.6}) as explained in the previous section. The rescaled saddle
setback $\Delta S_{resc}$ and the rescaled nut adjustment $\Delta N_{resc}$
from Eq. (\ref{eqn3.10}) were computed for each string, using the procedure
outlined in Sect. \ref{sect:compensation_model} and the geometrical and
physical parameters described above.

We then carefully measured the frequency of the sounds produced by pressing
each string onto the twenty frets of the fingerboard in the two possible
modes: without any compensation, i.e., setting the frets according to Eq.
(\ref{eqn2.4}), and with compensation, i.e., after shifting the position of
saddle and nut by the amounts specified in Table \ref{TableKey:table1} and
retuning the open string to the original note.

Table \ref{TableKey:table2} illustrates the frequency values for String 1,
obtained in the two different modes and compared to the theoretical values of
the same notes for the case of a \textquotedblleft perfect
intonation\textquotedblright\ of the instrument. The measurements were
repeated several times and the quantities in Table \ref{TableKey:table2}
represent average values.%

\begin{table}[tbp] \centering
\begin{tabular}
[c]{||l|l|l|l|l|l|l||}\hline\hline
$%
\begin{array}
[c]{c}%
\text{\textbf{String 1}}\\
\text{Fret}\\
\text{number}%
\end{array}
$ & Note & $%
\begin{array}
[c]{c}%
\text{\textbf{Perfect}}\\
\text{\textbf{intonation}}\\
\text{Frequency}\\
\text{(}Hz\text{)}%
\end{array}
$ & $%
\begin{array}
[c]{c}%
\text{\textbf{Without}}\\
\text{\textbf{compensation}}\\
\text{Frequency}\\
\text{(}Hz\text{)}%
\end{array}
$ & $%
\begin{array}
[c]{c}%
\text{\textbf{Without}}\\
\text{\textbf{compensation}}\\
\text{Freq. deviation}\\
\text{(}cents\text{)}%
\end{array}
$ & $%
\begin{array}
[c]{c}%
\text{\textbf{With}}\\
\text{\textbf{compensation}}\\
\text{Frequency}\\
\text{(}Hz\text{)}%
\end{array}
$ & $%
\begin{array}
[c]{c}%
\text{\textbf{With}}\\
\text{\textbf{compensation}}\\
\text{Freq. deviation}\\
\text{(}cents\text{)}%
\end{array}
$\\\hline
\multicolumn{1}{||c|}{$0$} & \multicolumn{1}{|c|}{$C_{3}$} &
\multicolumn{1}{|c|}{$130.813$} & \multicolumn{1}{|c|}{$130.813$} &
\multicolumn{1}{|c|}{$0$} & \multicolumn{1}{|c|}{$130.813$} &
\multicolumn{1}{|c||}{$0$}\\\hline
\multicolumn{1}{||c|}{$1$} & \multicolumn{1}{|c|}{$C_{3}^{\#}$} &
\multicolumn{1}{|c|}{$138.591$} & \multicolumn{1}{|c|}{$143.832$} &
\multicolumn{1}{|c|}{$64.3$} & \multicolumn{1}{|c|}{$137.958$} &
\multicolumn{1}{|c||}{$-7.9$}\\\hline
\multicolumn{1}{||c|}{$2$} & \multicolumn{1}{|c|}{$D_{3}$} &
\multicolumn{1}{|c|}{$146.832$} & \multicolumn{1}{|c|}{$150.551$} &
\multicolumn{1}{|c|}{$43.3$} & \multicolumn{1}{|c|}{$147.323$} &
\multicolumn{1}{|c||}{$5.8$}\\\hline
\multicolumn{1}{||c|}{$3$} & \multicolumn{1}{|c|}{$D_{3}^{\#}$} &
\multicolumn{1}{|c|}{$155.563$} & \multicolumn{1}{|c|}{$159.126$} &
\multicolumn{1}{|c|}{$39.2$} & \multicolumn{1}{|c|}{$155.363$} &
\multicolumn{1}{|c||}{$-2.2$}\\\hline
\multicolumn{1}{||c|}{$4$} & \multicolumn{1}{|c|}{$E_{3}$} &
\multicolumn{1}{|c|}{$164.814$} & \multicolumn{1}{|c|}{$168.407$} &
\multicolumn{1}{|c|}{$37.3$} & \multicolumn{1}{|c|}{$164.070$} &
\multicolumn{1}{|c||}{$-7.8$}\\\hline
\multicolumn{1}{||c|}{$5$} & \multicolumn{1}{|c|}{$F_{3}$} &
\multicolumn{1}{|c|}{$174.614$} & \multicolumn{1}{|c|}{$178.348$} &
\multicolumn{1}{|c|}{$36.6$} & \multicolumn{1}{|c|}{$173.933$} &
\multicolumn{1}{|c||}{$-6.8$}\\\hline
\multicolumn{1}{||c|}{$6$} & \multicolumn{1}{|c|}{$F_{3}^{\#}$} &
\multicolumn{1}{|c|}{$184.997$} & \multicolumn{1}{|c|}{$188.754$} &
\multicolumn{1}{|c|}{$34.8$} & \multicolumn{1}{|c|}{$184.763$} &
\multicolumn{1}{|c||}{$-2.2$}\\\hline
\multicolumn{1}{||c|}{$7$} & \multicolumn{1}{|c|}{$G_{3}$} &
\multicolumn{1}{|c|}{$195.998$} & \multicolumn{1}{|c|}{$200.386$} &
\multicolumn{1}{|c|}{$38.3$} & \multicolumn{1}{|c|}{$195.878$} &
\multicolumn{1}{|c||}{$-1.1$}\\\hline
\multicolumn{1}{||c|}{$8$} & \multicolumn{1}{|c|}{$G_{3}^{\#}$} &
\multicolumn{1}{|c|}{$207.652$} & \multicolumn{1}{|c|}{$212.105$} &
\multicolumn{1}{|c|}{$36.7$} & \multicolumn{1}{|c|}{$207.632$} &
\multicolumn{1}{|c||}{$-0.2$}\\\hline
\multicolumn{1}{||c|}{$9$} & \multicolumn{1}{|c|}{$A_{3}$} &
\multicolumn{1}{|c|}{$220.000$} & \multicolumn{1}{|c|}{$224.644$} &
\multicolumn{1}{|c|}{$36.2$} & \multicolumn{1}{|c|}{$220.081$} &
\multicolumn{1}{|c||}{$0.6$}\\\hline
\multicolumn{1}{||c|}{$10$} & \multicolumn{1}{|c|}{$A_{3}^{\#}$} &
\multicolumn{1}{|c|}{$233.082$} & \multicolumn{1}{|c|}{$237.495$} &
\multicolumn{1}{|c|}{$32.5$} & \multicolumn{1}{|c|}{$233.136$} &
\multicolumn{1}{|c||}{$0.4$}\\\hline
\multicolumn{1}{||c|}{$11$} & \multicolumn{1}{|c|}{$B_{3}$} &
\multicolumn{1}{|c|}{$246.942$} & \multicolumn{1}{|c|}{$252.345$} &
\multicolumn{1}{|c|}{$37.5$} & \multicolumn{1}{|c|}{$247.123$} &
\multicolumn{1}{|c||}{$1.3$}\\\hline
\multicolumn{1}{||c|}{$12$} & \multicolumn{1}{|c|}{$C_{4}$} &
\multicolumn{1}{|c|}{$261.626$} & \multicolumn{1}{|c|}{$266.338$} &
\multicolumn{1}{|c|}{$30.9$} & \multicolumn{1}{|c|}{$261.505$} &
\multicolumn{1}{|c||}{$-0.8$}\\\hline
\multicolumn{1}{||c|}{$13$} & \multicolumn{1}{|c|}{$C_{4}^{\#}$} &
\multicolumn{1}{|c|}{$277.183$} & \multicolumn{1}{|c|}{$281.958$} &
\multicolumn{1}{|c|}{$29.6$} & \multicolumn{1}{|c|}{$277.076$} &
\multicolumn{1}{|c||}{$-0.7$}\\\hline
\multicolumn{1}{||c|}{$14$} & \multicolumn{1}{|c|}{$D_{4}$} &
\multicolumn{1}{|c|}{$293.665$} & \multicolumn{1}{|c|}{$298.545$} &
\multicolumn{1}{|c|}{$28.5$} & \multicolumn{1}{|c|}{$293.688$} &
\multicolumn{1}{|c||}{$0.1$}\\\hline
\multicolumn{1}{||c|}{$15$} & \multicolumn{1}{|c|}{$D_{4}^{\#}$} &
\multicolumn{1}{|c|}{$311.127$} & \multicolumn{1}{|c|}{$315.276$} &
\multicolumn{1}{|c|}{$22.9$} & \multicolumn{1}{|c|}{$311.463$} &
\multicolumn{1}{|c||}{$1.9$}\\\hline
\multicolumn{1}{||c|}{$16$} & \multicolumn{1}{|c|}{$E_{4}$} &
\multicolumn{1}{|c|}{$329.628$} & \multicolumn{1}{|c|}{$334.822$} &
\multicolumn{1}{|c|}{$27.1$} & \multicolumn{1}{|c|}{$329.787$} &
\multicolumn{1}{|c||}{$0.8$}\\\hline
\multicolumn{1}{||c|}{$17$} & \multicolumn{1}{|c|}{$F_{4}$} &
\multicolumn{1}{|c|}{$349.228$} & \multicolumn{1}{|c|}{$353.408$} &
\multicolumn{1}{|c|}{$20.6$} & \multicolumn{1}{|c|}{$348.785$} &
\multicolumn{1}{|c||}{$-2.2$}\\\hline
\multicolumn{1}{||c|}{$18$} & \multicolumn{1}{|c|}{$F_{4}^{\#}$} &
\multicolumn{1}{|c|}{$369.994$} & \multicolumn{1}{|c|}{$373.545$} &
\multicolumn{1}{|c|}{$16.5$} & \multicolumn{1}{|c|}{$370.330$} &
\multicolumn{1}{|c||}{$1.6$}\\\hline
\multicolumn{1}{||c|}{$19$} & \multicolumn{1}{|c|}{$G_{4}$} &
\multicolumn{1}{|c|}{$391.996$} & \multicolumn{1}{|c|}{$396.597$} &
\multicolumn{1}{|c|}{$20.2$} & \multicolumn{1}{|c|}{$393.335$} &
\multicolumn{1}{|c||}{$5.9$}\\\hline
\multicolumn{1}{||c|}{$20$} & \multicolumn{1}{|c|}{$G_{4}^{\#}$} &
\multicolumn{1}{|c|}{$415.305$} & \multicolumn{1}{|c|}{$418.742$} &
\multicolumn{1}{|c|}{$14.3$} & \multicolumn{1}{|c|}{$417.068$} &
\multicolumn{1}{|c||}{$7.3$}\\\hline\hline
\end{tabular}
\caption{Frequency values of the different notes obtained with String 1: theoretical perfect intonation values are compared to the experimental values with and without compensation.
Also shown are the frequency deviations (in cents) from the theoretical values, for both cases.}\label{TableKey:table2}%
\end{table}%

In this table, fret number zero represents the open string being plucked, so
there is no difference in frequency for the three cases. On the contrary, for
all the other frets, the frequencies without compensation are considerably
higher than the theoretical values for a perfectly intonated instrument. This
results in the pitch\footnote{We note that the frequency of the sound produced
is the physical quantity we measured in our experiments. The pitch is defined
as a sensory characteristic arising out of frequency, but also affected by
other subjective factors which depend upon the individual. It is beyond the
scope of this paper to consider these additional subjective factors.} of these
notes to be perceived being higher (or sharper) than the correct pitch. In
fact, when we \textquotedblleft played\textquotedblright\ our monochord
sonometer in this first situation, it sounded definitely out of tune. The
frequency values obtained instead by using our compensation correction appear
to be much closer to the theoretical values, thus effectively improving the
overall intonation of our monochord instrument.

In Table \ref{TableKey:table2} we also show the frequency deviation of each
note from the theoretical value of perfect intonation for both cases: with and
without compensation. The frequency shifts are expressed in cents (see
definition in note at the end of Sect. \ref{description_apparatus}) rather
than in Hertz, since the former unit is more suitable to measure how the human
ear perceives different sounds to be in tune or out of tune. The frequency
deviation values illustrate more clearly the effectiveness of the compensation
procedure: without compensation the deviation from perfect intonation ranges
between $14.3$ and $64.3\ cents$, with compensation this range is reduced to
values between $-7.9$ and $+7.3\ cents$.

\begin{figure}
[ptb]
\begin{center}
\fbox{\ifcase\msipdfoutput
\includegraphics[
width=\textwidth
]%
{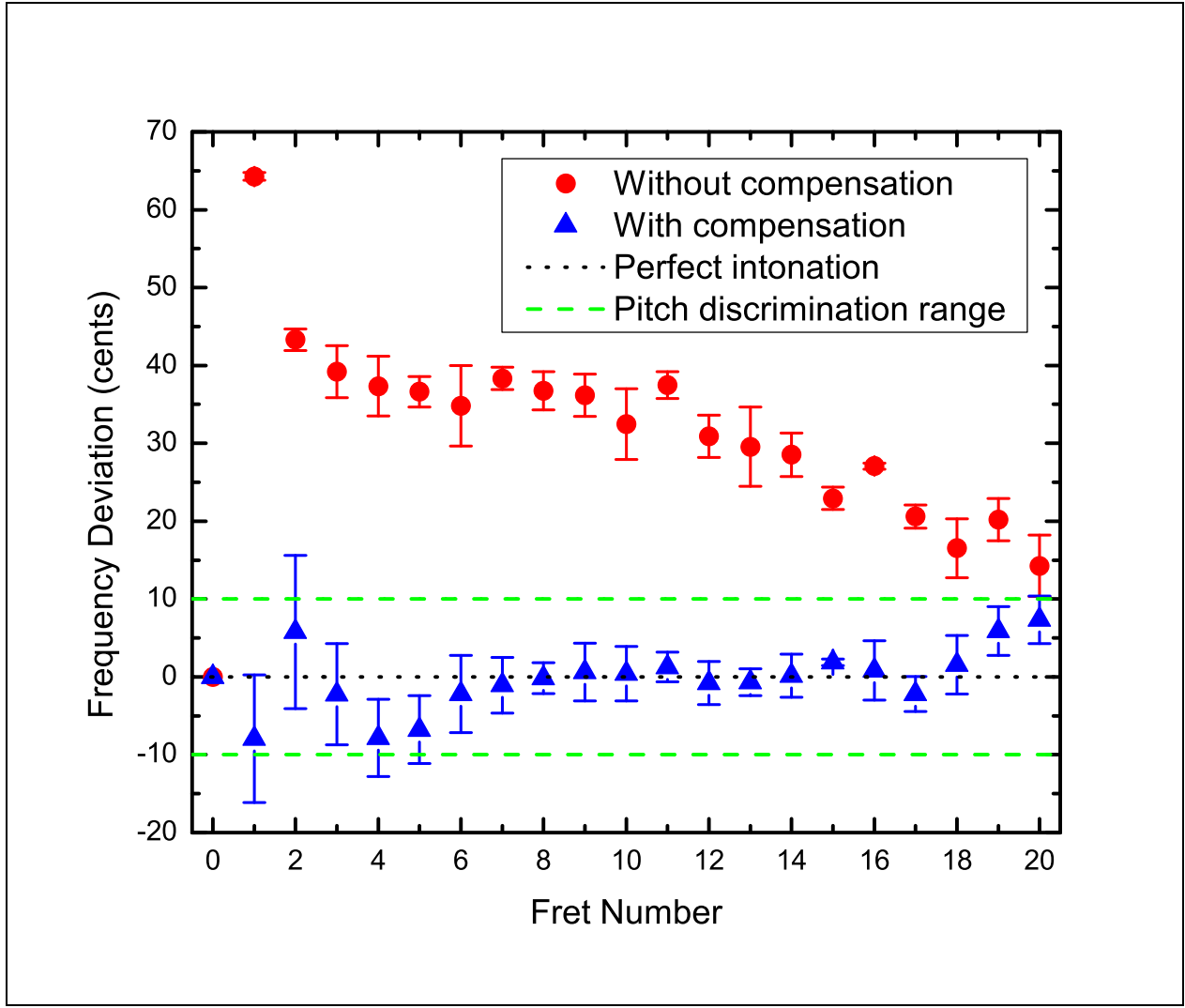}%
\else
\includegraphics[
width=\textwidth
]%
{C:/swp55/Docs/guitar1.1/AJP/arxiv_v2/graphics/Varieschi_GowerFig04__4.pdf}%
\fi
}\caption[Frequency deviation from perfect intonation level for notes obtained
with String 1.]{Frequency deviation from perfect intonation level (black
dotted line) for notes obtained with String 1. Red circles denote results
without compensation, while blue triangles denote results with compensation.
Also shown (region between green dashed lines) is the approximate pitch
discrimination range for frequencies related to this string. }%
\label{fig4}%
\end{center}
\end{figure}

In view of the previous discussion, we prefer to plot our results for String 1
in terms of the frequency deviation of each note from the theoretical value of
perfect intonation. Fig. \ref{fig4} shows these frequency deviations for each
fret number (corresponding to the different musical notes of Table
\ref{TableKey:table2}) in the two cases: without compensation (red circles)
and with compensation (blue triangles). Error bars are also included, coming
from the computed standard deviations of the measured frequency values.

We also show in the same figure the so-called pitch discrimination range
(region between green dashed lines), i.e., the difference in pitch which an
individual can effectively detect when hearing two different notes in rapid
succession. In other words, notes within this range will not be perceived as
different in pitch by human ears. It can be easily seen in Fig. \ref{fig4}
that all the (red) values without compensation are well outside the pitch
discrimination range, thus will be perceived as out of tune (in particular as
sharper sounds). On the contrary, the (blue) values with compensation are in
general within the green dashed discrimination range of about $\pm
10\ cents$\footnote{This discrimination range was estimated, for the
frequenqies of String 1, according to the discussion in Ref. \cite{Olson},
pages 248-252. In general, this range varies from about $\pm5\ cents$ -
$\pm10\ cents$ for frequencies between $1000-2000\ Hz$, to even larger values
of $\pm40\ cents$ - $\pm50\ cents$ at lower frequencies, between
$60-120\ Hz$.}. The compensation procedure has virtually made them equivalent
to the perfect intonation values (corresponding to the zero cent deviation -
perfect intonation level, black dotted line in Figure \ref{fig4}). We note
again that fret number zero simply corresponds to playing the open string note
which is always perfectly tuned, therefore the experimental points for this
fret do not show any frequency deviation.%

\begin{figure}
[ptb]
\begin{center}
\fbox{\ifcase\msipdfoutput
\includegraphics[
width=\textwidth
]%
{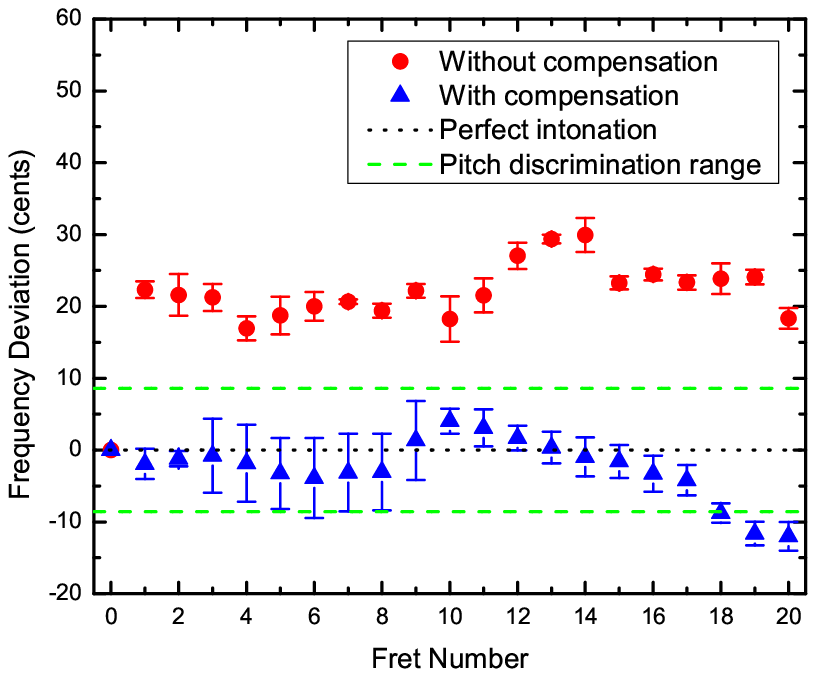}%
\else
\includegraphics[
width=\textwidth
]%
{C:/swp55/Docs/guitar1.1/AJP/arxiv_v2/graphics/Varieschi_GowerFig05__5.pdf}%
\fi
}\caption[Frequency deviation from perfect intonation level for notes obtained
with String 2.]{Frequency deviation from perfect intonation level (black
dotted line) for notes obtained with String 2. Red circles denote results
without compensation, while blue triangles denote results with compensation.
Also shown (region between green dashed lines) is the approximate pitch
discrimination range for frequencies related to this string. }%
\label{fig5}%
\end{center}
\end{figure}

\begin{figure}
[ptb]
\begin{center}
\fbox{\ifcase\msipdfoutput
\includegraphics[
width=\textwidth
]%
{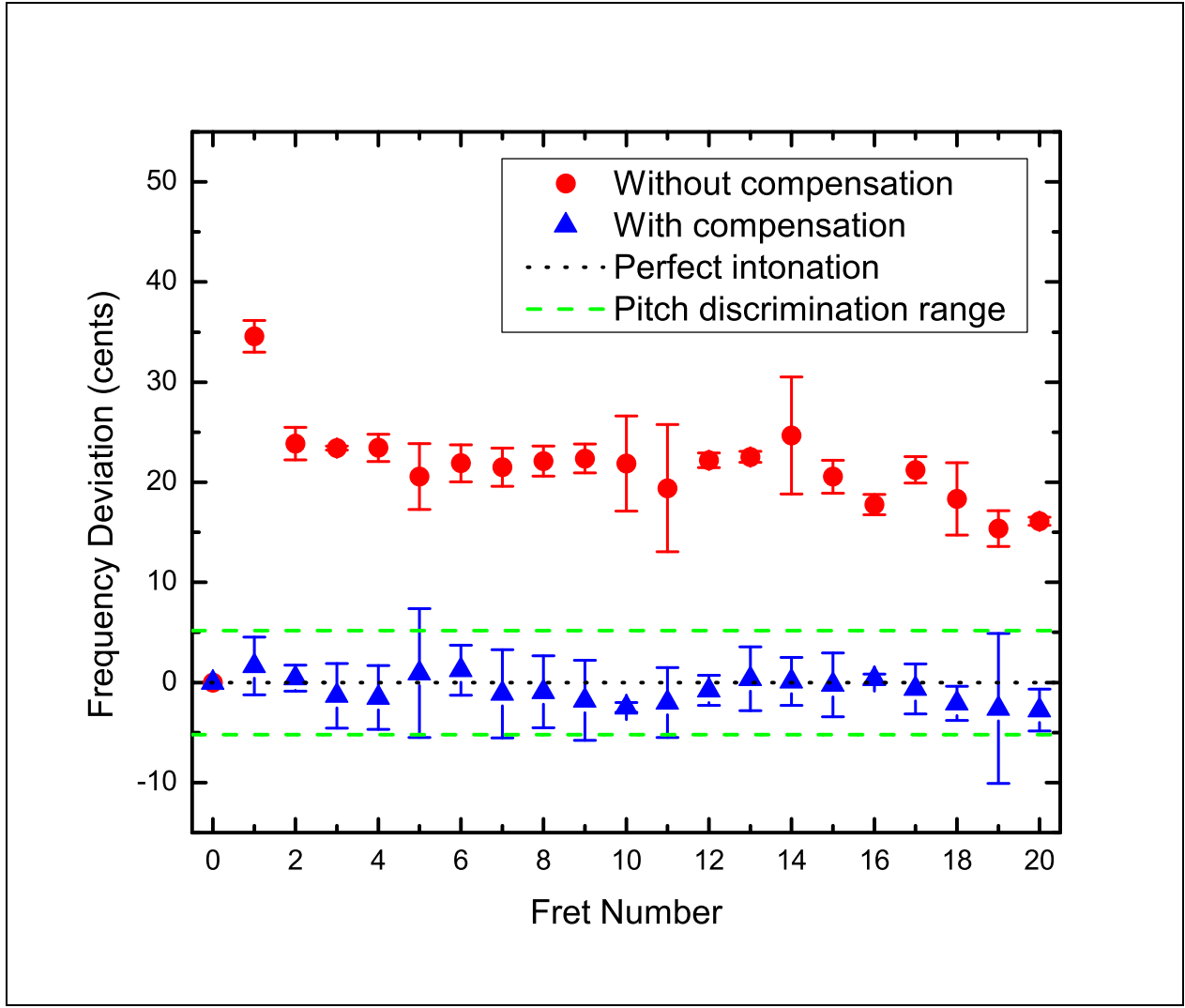}%
\else
\includegraphics[
width=\textwidth
]%
{C:/swp55/Docs/guitar1.1/AJP/arxiv_v2/graphics/Varieschi_GowerFig06__6.pdf}%
\fi
}\caption[Frequency deviation from perfect intonation level for notes obtained
with String 3.]{Frequency deviation from perfect intonation level (black
dotted line) for notes obtained with String 3. Red circles denote results
without compensation, while blue triangles denote results with compensation.
Also shown (region between green dashed lines) is the approximate pitch
discrimination range for frequencies related to this string. }%
\label{fig6}%
\end{center}
\end{figure}

We repeated the same type of measurements also for String 2 and 3, which were
tuned at higher frequencies as open strings (respectively as $F_{3}$ and
$C_{4}$, see Table \ref{TableKey:table1}). In this way we obtained
experimental sets of measured frequencies, with and without compensation, for
these two other strings, similar to those presented in Table
\ref{TableKey:table1}. For brevity, we will omit to report all these numerical
values, but we present in Figs. \ref{fig5} and \ref{fig6} the frequency
deviation plots, as we have done for String 1 in Fig. \ref{fig4}.%

The results in Figs. \ref{fig5} and \ref{fig6} are very similar to those in
the previous figure: the frequencies without compensation are much higher than
the perfect intonation level, while the compensation procedure is able to
reduce almost all the frequency values to the region within the green dashed
curves (the pitch discrimination range). Using again the procedure outlined in
Ref. \cite{Olson}, the discrimination ranges in Figs. \ref{fig5} and
\ref{fig6} were computed respectively as $\pm8.6\ cents$ and $\pm5.2\ cents$,
due to the different frequencies produced by these two other strings.

For the three cases we analyzed, we can conclude that the compensation
procedure described in this paper is very effective in improving the
intonation of each of the strings we used. Although more work on the subject
is needed (in particular we need to test also nylon strings, which are more
commonly used in classical guitars), we have proven that the intonation
problem of fretted string instruments can be analyzed and solved using
physical and mathematical models, which are more reliable than the empirical
methods developed by luthiers during the historical development of these instruments.%

\section{\label{sect:conclusions}Conclusions}

In this work we studied the mathematical models and the physics related to the
problem of intonation and compensation of fretted string instruments. While
this problem is usually solved in an empirical way by luthiers and instrument
makers, we have shown that it is possible to find a mathematical solution,
improving the original work by G. Byers and others, and that this procedure
can be effectively implemented in practical situations.

We have demonstrated how to use simple lab equipment, such as standard
sonometers and frequency measurement devices, to study the sounds produced by
plucked strings, when they are pressed onto a guitar-like fingerboard, thus
confirming the mathematical models for intonation and compensation. These
activities can also be easily presented in standard musical acoustics courses,
or used in sound and waves labs as an interesting variation of experiments
usually performed with classic sonometers.

\begin{acknowledgments}
This work was supported by a grant from the Frank R. Seaver College of Science
and Engineering, Loyola Marymount University. The authors would like to
acknowledge useful discussions with John Silva of Trilogy Guitars and with
luthier Michael Peters, who also helped us with the guitar fingerboard. We
thank Jeff Cady for his technical support and help with our experimental
apparatus. We are also very thankful to luthier Dr. Gregory Byers for sharing
with us important details of his original study on the subject and other
suggestions. Finally, the authors gratefully acknowledge the anonymous
reviewers for their useful comments and suggestions.
\end{acknowledgments}

\bibliographystyle{apsrev}
\bibliography{GUITAR1}

\end{document}